\documentstyle[12pt,epsfig]{article}

\begin{document}

\begin{flushright}
{\large INR 0945c/99}
\end{flushright}

\bigskip

\begin{center}{\Large Observability and Probability of Discovery
in Future Experiments}
\end{center}

\bigskip

  \begin{center}
{\large
    S.I.~Bityukov$^{1,2}$ , N.V.~Krasnikov$^3$ \\

Institute for Nuclear Research, Moscow, Russia
}
\end {center}

\vspace{1cm}

\begin{abstract}
We propose a method to estimate the probability of new physics
discovery in future high energy physics experiments. Physics
simulation gives both the average numbers $<N_b>$ of background 
and $<N_s>$ of signal events. We find that the proper definition
of the significance is
$S_{12} = \sqrt{<N_s>+<N_b>} - \sqrt{<N_b>}$ in comparison with often
used significances $S_1 = \displaystyle \frac{<N_s>}{\sqrt{<N_b>}}$ and
$S_2 = \displaystyle \frac{<N_s>}{\sqrt{<N_s> + <N_b>}}$.
We propose a method for taking into account systematic uncertainties
related to nonexact knowledge of background and signal cross sections.
An account of such systematics is very essential in the search for 
supersymmetry at LHC. We propose a method for estimation of exclusion 
limits on new physics in future experiments. We also estimate the
probability of new physics discovery in future experiments taking into
account systematical errors.
\end{abstract}

\vspace{1cm}
\small
\noindent
\rule{3cm}{0.5pt}\\
$^1$Institute for High Energy Physics, Protvino, Moscow region, Russia\\
$^2$E-mails: bityukov@mx.ihep.su,~~Serguei.Bitioukov$@$cern.ch\\
$^3$E-mails: krasniko@ms2.inr.ac.ru,~~Nikolai.Krasnikov$@$cern.ch\\
\vspace{1cm}

\section{ Introduction}

One of the common goals in the forthcoming experiments is the search for new 
phenomena. In the forthcoming high energy physics experiments 
(LHC, TEV22, NLC, ...) the main goal is the search for physics
beyond the Standard Model (supersymmetry, $Z'$-, $W'$-bosons, ...) and
the Higgs boson discovery as a final confirmation of the Standard Model.
In estimation of the discovery potential of the future experiments
(to be specific in this paper we shall use as an example CMS experiment
at LHC~\cite{1}) the background cross section 
is calculated and for the given integrated luminosity $L$ the average 
number of background events is $<N_b> = \sigma_b \cdot L$.
Suppose the existence of a new physics 
leads to the nonzero signal cross section $\sigma_s$ with the same 
signature as for the background cross section  that 
results in  the prediction
of the additional average number of signal events 
$<N_s> = \sigma_s \cdot L$ for the integrated luminosity $L$.

The total average number of the events is 
$<N_{ev}> = <N_s> + <N_b> = (\sigma_s + \sigma_b) \cdot L$.
So, as a result of new physics existence, we expect an excess 
of the average number of events. In real experiments the probability
of the realization of $n$ events is described by Poisson 
distribution~\cite{2}

\begin{equation}
f(n,<n>) = \frac{<n>^n}{n!} e^{-<n>}.
\end{equation}
Here $<n>$ is the average number of events.

Remember that the Poisson distribution $f(n,<n>)$ gives~\cite{3} the 
probability of finding exactly $n$ events in the given interval
of (e.g. space and time) when the events occur independently
of one another at an average rate  of $<n>$ per the 
given interval. For the Poisson distribution the variance $\sigma^2$
equals to $<n>$. So, to estimate the probability of the new physics discovery 
we have to compare the Poisson statistics with $<n> = <N_b>$ and
$<n> = <N_b> + <N_s>$. Usually, high energy physicists use the following  
``significances'' for testing the possibility to discover new physics
in an  experiment:

\begin{itemize}
\item[(a)]
``significance'' $S_1 = \displaystyle \frac{<N_s>}{\sqrt{<N_b>}}$~\cite{4},
\item[(b)]
``significance'' $S_2 = \displaystyle \frac{<N_s>}{\sqrt{<N_s> + <N_b>}}$~\cite{5,6}.
\end{itemize}

\noindent
A conventional claim is that for $S_1~(S_2) \ge 5$ we shall discover
new physics (here, of course, the systematical errors are ignored).
For $N_b \gg N_s$ the significances  $S_1$ and $S_2$ coincide (the 
search for Higgs boson through the $H \rightarrow \gamma \gamma$
signature). For the case when $N_s \sim N_b$,~$S_1$ and $S_2$ differ.
Therefore, a natural question arises: what is the correct definition for the 
significance $S_1$, $S_2$ or anything else ?

It should be noted  that there is a crucial difference between ``future''
experiment and the ``real'' experiment. In the ``real'' experiment the total 
number of events $N_{ev}$ is a given number (already has been measured)
and we compare it with $<N_b>$ when we test the validity of the
standard physics. So, the number of possible signal events is determined
as $N_s = N_{ev} - <N_b>$ and it is compared  with the average number of 
background events $<N_b>$. The fluctuation of the background is 
$\sigma_{fb} = \sqrt{N_b}$, therefore, we come to the $S_1$ significance
as the measure of the distinction from the standard physics.
In the conditions of the ``future'' experiment when we want to search
for new physics, we know only the average number of the background
events and the average number of the signal events, so we
have to compare the Poisson distributions $P(n,<N_b>)$ and
$P(n,<N_b>+<N_s>)$ to determine the probability to find new physics
in the future experiment.

In this paper we estimate the probability to discover new physics
in future experiments. We show that the
proper determination of the significance is 
$S = \sqrt{<N_s>+<N_b>} - \sqrt{<N_b>}$. We suggest a method
which takes into account systematic errors related to  nonexact
knowledge of the signal and background cross sections. We also propose
a method for the estimation of exclusion limits on new physics
in future experiments. Some of presented results has been published
in our early paper~\cite{8}.

The organization of the paper is the following. 
In the next Section we give a method
for the determination of the probability to find new physics in the future
experiment and calculate the probability to discover new physics
for the given $(<N_b>,~<N_s>)$ numbers of background and signal events
under the assumption that there are no systematic errors. In Section 3 
we estimate the influence of the systematics related to nonexact 
knowledge of the signal and background cross sections on the probability 
to discover new physics in future experiments. In Section 4 we describe 
a method for the estimation of exclusion limits on new physics
in future experiments. 
In Section 5 we estimate the probability of new physics discovery 
in future experiments.
Section 6 contains concluding remarks.

\section{An analysis of statistical fluctuations}

Suppose that for some future experiment we know the average number of 
the background and signal events $<N_b>,~<N_s>$. As it has been mentioned
in the Introduction, the probability of realization of $n$ events in an 
experiment is given by the Poisson distribution

\begin{equation}
P(n,<n>) = \frac{<n>^n}{n!} e^{-<n>},
\end{equation}
where $<n> = <N_b>$ for the case of the absence of new physics
and $<n> = <N_b> + <N_s>$ for the case when new physics exists.
So, to determine the probability to discover new physics in future
experiment, we have to compare the Poisson distributions with
$<n> = <N_b>$ (standard physics) and $<n> = <N_b> + <N_s>$ (new physics).

Consider, at first, the case when $<N_b>~~\gg 1$, $<N_s>~~\gg 1$.
In this case the Poisson distributions approach the Gaussian 
distributions~\footnote{With a precision defined by the tails 
(see Section 5).}

\begin{equation}
P_G(n,\mu,\sigma^2) = \frac{1}{\sigma \sqrt{2 \pi}} \cdot
e^{-\frac{(n - \mu)^2}{2 \sigma^2}},
\end{equation}
with $\mu = \sigma^2$ and $\mu = <N_b>$ or $\mu = <N_b> + <N_s>$. 
Here $n$ is a real number. Note that for the Poisson distribution
the mean equals to the variance.

The Gaussian distribution describes  the probability density to realize
$n$ events in the future experiment provided the average number of events
$<n>$ is a given number. In Fig.1 we show two Gaussian distributions 
$P_G$ with $<n> = <N_b> = 53$ and $<n> = <N_b> + <N_s>$ = 104
(\cite{6}, Table.13, cut 6). As is clear from Fig.1 
the common area for these two curves (the first curve shows the
``standard physics'' events distribution and the second one 
gives the ``new physics'' events distribution) is the probability 
that ``new physics'' can be described by the ``standard physics''.
In other words, suppose we know for sure that new physics takes place
and the probability density of the events realization is described
by curve II ($f_2(x) = P_G(x,<N_b>+<N_s>,<N_b>+N_s>))$.
The probability $\kappa$ that the ``standard physics''
(curve I ($f_1(x) = P_G(x,<N_b>,<N_b>) $)) can imitate  new physics 
(i.e. the probability that we measure ``new physics''
but we think that it is described by the ``standard physics'')
is described by common area of curve I and II. 

\begin{figure}[htpb]
  \begin{center}
    \resizebox{7cm}{!}{\includegraphics{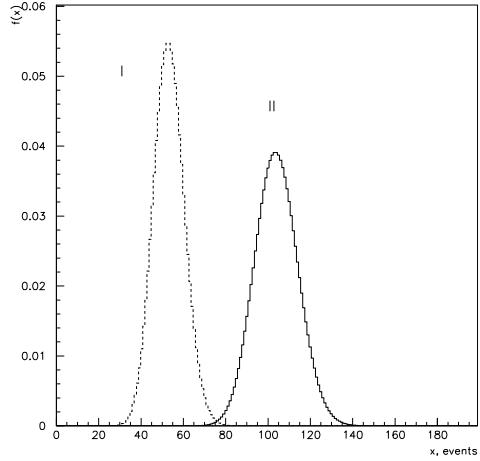}} 
\caption{The probability density functions
$f_{1,2}(x) \equiv P_G(x,\mu_{1,2},\sigma^2)$ for $\mu_1 = <N_b> = 53$ 
and $\mu_2 = <N_b> + <N_s> = 104$.}
    \label{fig:1} 
  \end{center}
\end{figure}
        
Numerically, we find that 

\begin{eqnarray}
\kappa &=& \frac{1}{\sqrt{2 \pi} \sigma_2}
\int_{-\infty}^{\nonumber \sigma_1 \sigma_2}exp[-\frac{(x-\sigma_2^2)^2}{2 \sigma_2^2}]dx +
\frac{1}{\sqrt{2 \pi} \sigma_1}
\int_{\sigma_1 \sigma_2}^{\infty}exp[-\frac{(x-\sigma_1^2)^2}{2 \sigma_1^2}]dx\\  
&=&\frac{1}{\sqrt{2 \pi}}
[\int_{-\infty}^{\sigma_1-\sigma_2}exp[-\frac{y^2}{2}]dy +
\int_{\sigma_2-\sigma_1}^{\infty}exp[-\frac{y^2}{2}]dy] \\ \nonumber
&=&1 - erf(\frac{\sigma_2-\sigma_1}{\sqrt{2}}).
\end{eqnarray}
Here $\sigma_1 = \sqrt{N_b}$ and $\sigma_2 = \sqrt{N_b + N_s}$.
The transformation of the distributions to standard normal distribution
and the exploitation of the equality
\begin{center}
$\displaystyle
\frac{x_0 - \sigma_1^2}{\sigma_1} = - \frac{x_0 - \sigma_2^2}{\sigma_2}$
\end{center}
allows one to find the point $x_0$ of the intersection of the curves I and II.

Let us discuss the meaning of our definition (4). For $x \leq x_0 = \sigma_{1}
\sigma_{2}$ we have $f_1(x) \geq f_2(x)$, i.e. the probability density of the 
standard physics realization is higher than the probability density of 
new physics realization. Therefore for $x \leq x_0$ we do not have any 
indication in favour of new physics. The probability that the number of 
events is less than $x_0$ is 
$\displaystyle \alpha = \int_{-\infty}^{x_0} f_2(x)dx$. 
For $x > x_0$ $f_2(x) > f_1(x)$ that gives evidence in favour of new physics 
existence. However the probability of the background events with 
$x > x_0$  is different from zero and is equal to 
$\displaystyle \beta = \int _{x_0}^{\infty}f_1(x)dx$. 
So we have two types of the errors. 
For $x  \leq x_0$ we do not have any evidence in favour of new physics 
(even in this case the probabilty of new physics realization 
is different from zero). For $x >x_0$ we have evidence in favour of new 
physics. However for $x > x_0$ the fluctuations of the background can 
imitate new physics. So the probability that standard physics can 
imitate new physics has two components $\alpha$ and $\beta$ and it is 
equal to $\kappa = \alpha + \beta$. 
If $\kappa$ equals to $1$ new physics will
never be found in the experiment, if $\kappa$ equals to $0$
the first measurement with probability one has to answer 
the question about presence or absence of new physics  (this case
is not realized for Poisson distribution). 
In other words one can say that
the area of intersection of the probability density functions of the
pure background and the background plus signal 
is the measure  of the future experiment undiscovery potential.

As follows from formula (4) the role of the significance $S$
plays 

\begin{equation}
S_{12} = \sigma_2 - \sigma_1 = \sqrt{N_b + N_s} - \sqrt{N_b}. 
\end{equation}

Note that  in refs.\cite{7}  the following criterion of the signal
discovery has been used. The signal was assumed to be observable
if $(1 - \epsilon) \cdot 100\%$ upper confidence level for the
background event rate is equal to $(1 - \epsilon) \cdot 100\%$
lower confidence level for background plus signal
$(\epsilon = 0.01 - 0.05)$. The corresponding
significance is similar to our significance $S_{12}$. The
difference is that in our approach the probability
$\kappa$ that new physics is described by standard physics is
equal to $2 \epsilon$. 

It means that for 
$S_{12} = 1, 2, 3, 4, 5, 6$ the probability $\kappa$ is correspondingly
$\kappa = 0.31, 0.046, 0.0027, 6.3 \cdot 10~^{-5}$, $5.7 \cdot~10^{-7},
2.0 \cdot~10^{-9}$ in accordance with a general picture. 
As it has been mentioned in the Introduction two definitions of the significance
are mainly used in the literature: 
 $S_1 = \displaystyle \frac{<N_s>}{\sqrt{<N_b>}}$~\cite{4} and
 $S_2 = \displaystyle \frac{<N_s>}{\sqrt{<N_s> + <N_b>}}$~\cite{5}.
The significance $S_{12}$ is expressed in terms of the significances
$S_1$ and $S_2$ as $S_{12} = \displaystyle \frac{S_1 S_2}{S_1 + S_2}$.

For $<N_b> \gg <N_s>$ (the search for Higgs boson through 
$H \rightarrow \gamma \gamma$ decay mode) we find that 

\begin{equation}
S_{12} \approx 0.5~S_1 \approx 0.5~S_2.
\end{equation}
It means that for $S_1 = 5$ (according to a common convention the
$5 \sigma$ confidence level means a new physics discovery) the
real significance is $S_{12} = 2.5$, that  corresponds 
to $\kappa = 1.24 \%$~(Fig.2).

\begin{figure}[htpb]
  \begin{center}
    \resizebox{7cm}{!}{\includegraphics{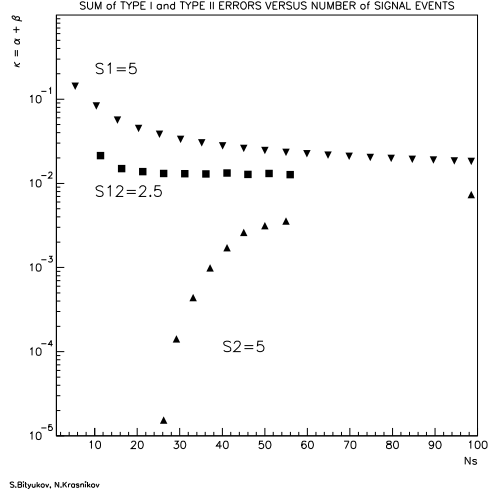}} 
\caption{The dependence of $\kappa$ on number of signal events
for cases $S_1=5,~S_2=5$ and $S_{12}=2.5$.}
    \label{fig:2} 
  \end{center}
\end{figure}

For the case $N_s = k N_b$, $S_{12} = k_{12} S_2$, where for
$k = 0.5, 1, 4, 10$ the values of $k_{12}$ are 
$k_{12} = 0.55, 0.59, 0.69, 0.77$, correspondingly. For not too high values of 
$<N_b>$ and $<N_b> + <N_s>$, we have to compare the Poisson distributions
directly. Again for the Poisson distribution $P(n,<n>)$ with the area
of definition for nonnegative integers we can define
$P(x,<n>)$ for real $x$ as 

\begin{equation}
\displaystyle \tilde P(x, <n>) = \left\{
\begin{array}{ll}
               0,        & x < 0, \\
     P([x], <n>),        & x \ge 0.    
\end{array}
\right.
\end{equation}
It is evident that 

\begin{equation}
\int_{-\infty}^{\infty}{\tilde P(x, <n>) dx} = 1.
\end{equation}

So, the generalization of the previous determination of $\kappa$
in our case is straightforward, namely, $\kappa$ is nothing but 
the common area of the curves described by $\tilde P(x, <N_b>)$
(curve I) and $\tilde P(x, <N_b> + <N_s>)$ (curve II) (see, Fig.3).

\begin{figure}[htpb]
  \begin{center}
    \resizebox{7cm}{!}{\includegraphics{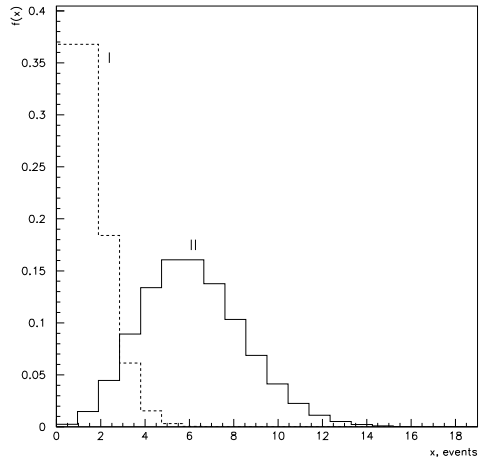}} 
\caption{The probability density functions
$f_{1,2}(x) \equiv \tilde P(x,\mu_{1,2})$ for $\mu_1$ = $<N_b>$ = $1$ 
and $\mu_2 = <N_b> + <N_s> = 6$.}
    \label{fig:3} 
  \end{center}
\end{figure}
        
\noindent
One can find that 
$ \kappa = \alpha + \beta$, where

%
%

$\displaystyle \alpha = \sum^{n_0}_{n=0}\frac{(<N_b> + <N_s>)^n}{n!}
\displaystyle e^{-(<N_b> + <N_s>)} =$  

\hfill              $1 - F(2<N_b>+2<N_s>|2n_0+2)$,

$\displaystyle \beta =  \sum_{n = n_0 + 1}^{\infty}\frac{(<N_b>)^n}{n!}
\displaystyle e^{-<N_b>} = F(2<N_b>|2n_0+2)$,

$F(\chi^2|n)$ = 
$\displaystyle \frac{1}{2^{n \over 2} \Gamma({n \over 2})}
\int^{\chi^2}_0{e^{-{t \over 2}} t^{{n \over 2} - 1} dt}$ 
(see, for example, \cite{9}) and 
$\displaystyle n_0 = [ \frac{<N_s>}{ln(1 + \frac{<N_s>}{<N_b>})}].$

\noindent
Numerical results  are presented  in Tables 1-6. 

As it follows
from these Tables for finite values of $<N_s>$ and $<N_b>$
the deviation from asymptotic formula (4) is essential.
For instance, for $N_s = 5$, $N_b = 1~~(S_1 = 5)$~
$\kappa = 14.2\%$. For $N_s = N_b = 25~~(S_1 = 5)$~
$\kappa = 3.8\%$, whereas asymptotically for $N_s \gg 1$ we find 
$\kappa = 1.24\%$. 
Similar situation takes place for $N_s \sim N_b$.

\section{An account of systematic errors related to nonexact
knowledge of background and signal cross sections}

In the previous section we determined the statistical error $\kappa$
(the probability that ``new physics'' is described by ``standard physics'').
In this section we investigate the influence of the systematical errors
related to a nonexact knowledge of the background and signal 
cross sections on the probability $\kappa$ not to confuse a new
physics with the old one.

Denote the Born background and signal cross sections as 
$\sigma_b^0$ and $\sigma_s^0$. An account of one loop corrections leads to
$\sigma_b^0 \rightarrow \sigma_b^0(1+\delta_{1b})$ and   
$\sigma_s^0 \rightarrow \sigma_s^0(1+\delta_{1s})$, where typically 
$\delta_{1b}$ and $\delta_{1s}$ are {\cal O}(0.5).

Two loop corrections at present are not known. So, we can assume that the 
uncertainty related with nonexact knowledge of cross sections is around
$\delta_{1b}$ and $\delta_{1s}$ correspondingly. In other words, we assume that
the exact cross sections lie in the intervals 
$(\sigma_b^0, \sigma_b^0 (1+2 \delta_{1b}))$ and  
$(\sigma_s^0, \sigma_s^0 (1+2 \delta_{1s}))$.
The average number of background and signal events lie in the intervals  

\begin{equation}
(<N_b^0>, <N_b^0> (1+2 \delta_{1b})) 
\end{equation}
and  

\begin{equation}
(<N_s^0>, <N_s^0> (1+2 \delta_{1s})), 
\end{equation}
where $<N_b^0> = \sigma_b^0 \cdot L$, $<N_s^0> = \sigma_s^0 \cdot L$. 

To determine the probability that the new physics is described by the 
old one, we 
again have to compare  two Poisson distributions with and without new physics
but in distinction from Section 2 we have to compare the Poisson distributions 
in which the average numbers lie in some intervals. So, a priori the only
thing we know is that the average numbers of background and signal events
lie in the intervals (9) and (10), but we do not know the exact values of $<N_b>$
and $<N_s>$. To determine the probability that 
the new physics is described by the old one,
consider the worst case~\footnote{There is a problem to determine
systematic uncertainty probability distributions for theoretical predictions
under consideration.} 
when we think that new  physics 
is described by the minimal number of average events 

\begin{equation}
<N_b^{min}> = <N_b^0> + <N_s^0>.
\end{equation}

Due to the fact that we do not know the exact value of the background
cross section,  consider the worst case when the average 
number of background events is equal to $<N_b^0> (1 + 2 \delta_{1b})$.
So, we have to compare the Poisson distributions with 
$<n> = <N_b^0> + <N_s^0> =$

$<N_b^0> (1 + 2 \delta_{1b}) + (<N_s^0>  - 2 \delta_{1b} <N_b^0>)$
and $<n> = <N_b^0> (1 + 2 \delta_{1b})$.
Using the result of the previous Section, we find that for case  
$<N_b^0>~\gg~1, <N_s^0>~\gg~1$  the effective significance is

\begin{equation}
S_{12s} = \sqrt{<N_b^0>+<N_s^0>} - \sqrt{<N_b^0>(1+2\delta_{1b})}.
\end{equation}

For the limiting case $\delta_{1b} \rightarrow 0$, we reproduce
formula (5). For not too high values of $<N_b^0>$ and $<N_s^0>$, we
have to use the results of the previous section (Tables 1-6).

As an example consider the case when $\delta_{1b} = 0.5$,
$<N_s> = 100$, $<N_b> = 50$ (typical situation for sleptons search). 
In this case we find that\\

 $S_1 = \displaystyle \frac{<N_s>}{\sqrt{<N_b>}} = 14.1,$

 $S_2 = \displaystyle \frac{<N_s>}{\sqrt{<N_s> + <N_b>}} = 8.2$

 $S_{12} = \sqrt{<N_b>+<N_s>} - \sqrt{<N_b>} = 5.2,$

 $S_{12s} = \sqrt{<N_b>+<N_s>} - \sqrt{2<N_b>} = 2.25.$\\

The difference between CMS adopted significance
$S_2~=~8.2$ (that corresponds to the probability 
$\kappa = 0.24 \cdot 10^{-15}$)
and the significance $S_{12s}~=~2.25$ taking into account 
systematics related to  nonexact knowledge of background cross section 
is factor 3.6. 
The direct comparison of the Poisson distributions 
with $<N_b>(1 + 2\delta_{1b}) = 100$ and $<N_b>(1 + 2\delta_{1b})
 + <N_{s,eff}>$ ( $<N_{s,eff}> = <N_s> - 2\delta_{1b}<N_b> = 50)$ 
gives $\kappa_{s} 
= 0.0245$.

Another example is with $<N_s> = 28$, $<N_b>=8$ and $\delta_{1b} = 0.5$. 
For such example we have $S_1 = 9.9$, $S_2 = 4.7$, $S_{12} = 3.2$, 
$S_{12s} = 2.0$, $\kappa_{s} =0.045$.

So, we see that an account of the systematics related to nonexact knowledge 
of background cross sections is very essential and it decreases the LHC SUSY 
discovery potential.

\section{Estimation of exclusion limits on new physics}

In this section we generalize the results of the previous sections
to obtain exclusion limits on signal cross section (new physics).

Suppose we know the background cross section $\sigma_b$ and we
want to obtain bound on signal cross section $\sigma_s$ which depends on
some parameters (masses of new particles, coupling constants, ...)
and describes some new physics beyond standard model. Again as in 
Section 2 we have to compare two Poisson distributions with and 
without new physics. The results of Section 2 are trivially generalized
for the case of the estimation of exclusion limits on signal 
cross section and, hence, on parameters (masses, coupling constants, ...)
of new physics.

Consider at first the case when $<N_b> = \sigma_b \cdot L \gg 1$,
$<N_s> = \sigma_s \cdot L \gg 1$ and the Poisson distributions
approach the Gaussian distributions. As it has been mentioned in
Section 2 the common area of the Gaussian curves with background events
and with background plus signal events is the probability that 
"new physics" can be described by the "standard physics".
For instance, when we require the probability that "new physics"
can be described by the "standard physics" is less or equal $10\%$
($S_{12}$ in formula (5) is larger than 1.64) it means that the
formula  

\begin{equation}
\sqrt{<N_b>+<N_s>} - \sqrt{<N_b>} \le 1.64
\end{equation}
gives us $90\%$ exclusion limit on the average number of signal events
$<N_s>$. In general case when we require the probability that 
"new physics" can be described by the "standard physics" is more
or less $\epsilon$ the formula

\begin{equation}
\sqrt{<N_b>+<N_s>} - \sqrt{<N_b>} \le S(\epsilon)
\end{equation}
allows us to obtain $1-\epsilon$ exclusion limit on signal cross section.
Here $S(\epsilon)$ is determined 
by the formula~(4)~\footnote{Note that $S(1\%) = 2.57$,
$S(2\%) = 2.33$, $S(5\%) = 1.96$ and $S(10\%) = 1.64$}, i.e. we suppose that 
$\epsilon = \kappa$. 
It should be stressed
that in fact the requirement that "new physics" with the probability
more or equal to $\epsilon$ can be described by the "standard physics"
is our definition of the exclusion limit as $(1-\epsilon)$ probability
for signal cross section. From the formula (14) we find that

\begin{equation}
\sigma_s \le \displaystyle \frac{S^2(\epsilon)}{L} + 
2 S(\epsilon) \sqrt{\frac{\sigma_b}{L}}.\\
\end{equation}

For the case of not large values of $<N_b>$ and $<N_s>$ we have to compare
the Poisson distributions directly and the corresponding method
has been formulated in Section 2. As an example in Table 7 we give 
$90\%$ exclusion limits on the signal cross section for 
$L = 10^4pb^{-1}$ and for different values of background cross sections.

Formulae (14), (15) do not take into account the influence of the
systematical errors related to nonexact knowledge of the 
background cross sections on the exclusion limits for signal cross section.
To take into account such systematics we have to use the results of
Section 3. The corresponding generalization of the formulae (14) and (15)
is straightforward, namely:

\begin{equation}
\sqrt{<N_b>+<N_s>} - \sqrt{<N_b>(1+2\delta_{1b})} \le S(\epsilon),
\end{equation}

\begin{equation}
\sigma_s \le \displaystyle \frac{S^2(\epsilon)}{L} + 
2 S(\epsilon) \displaystyle \sqrt{\frac{\sigma_b(1+2\delta_{1b})}{L}} + 
2 \delta_{1b} \sigma_b.
\end{equation}

Remember that $\delta_{1b}$ describes theoretical uncertainty in the
calculation of the background cross section. As an example, in
Table 8 we give $90\%$ exclusion limits on the signal cross section
for $L = 10^4pb^{-1},~~2 \delta_{1b} = 0.25$ and for different values
of background cross sections.

Note that in refs.\cite{9,10} different and strictly speaking "ad hoc" methods
to derive exclusion limits in future experiments has been suggested. 
As is seen from Fig.4 the essential differences 
in values of the exclusion limits take place.
Let us compare these methods by the use of the equal probability test~\cite{11}.
\begin{figure}[htpb]
  \begin{center}
    \resizebox{7cm}{!}{\includegraphics{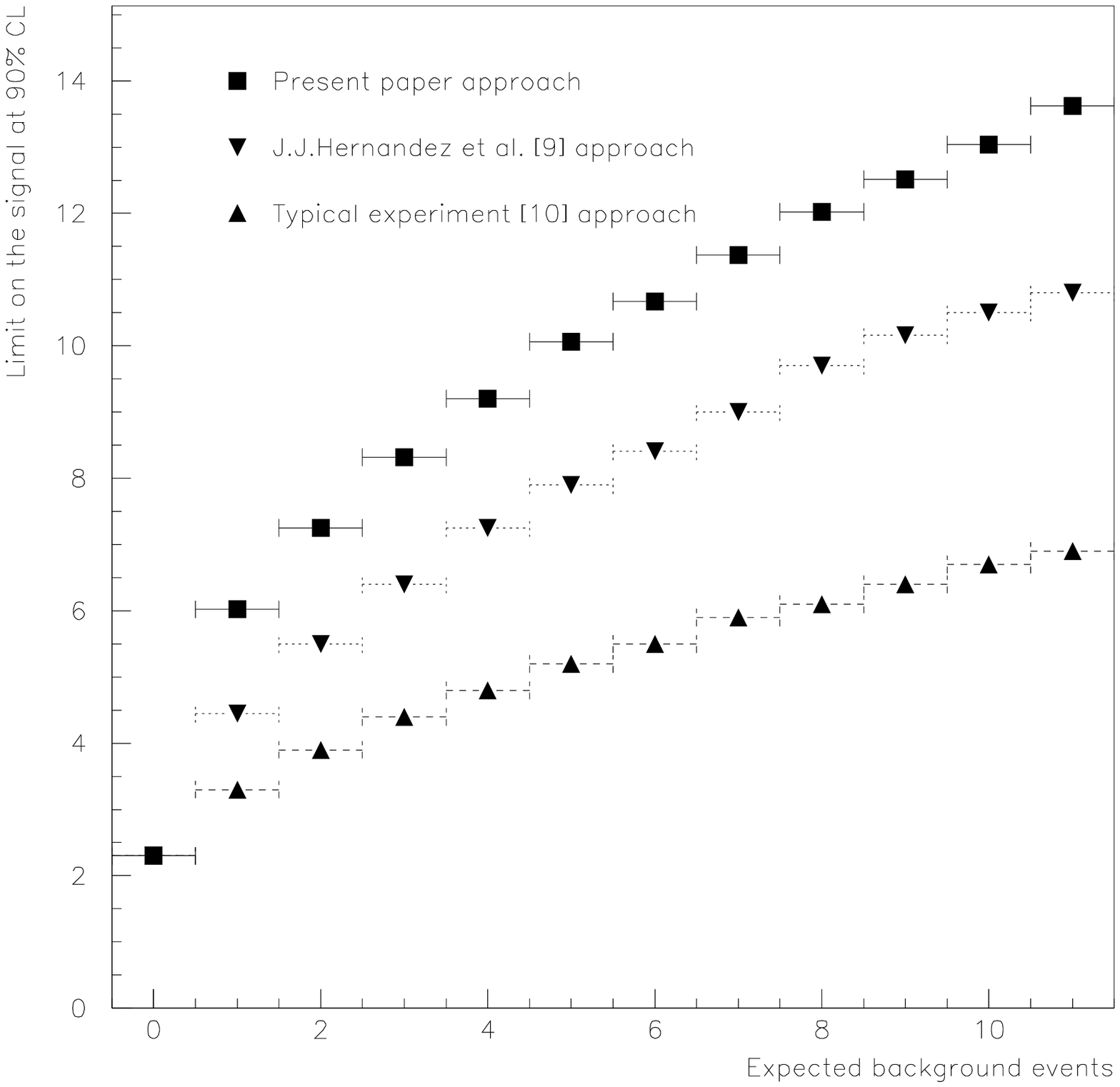}} 
\caption{Estimations of the 90\%~CL upper limit on the signal in a future 
experiment as a function of the expected background. The method proposed
in ref.~[10] gives the values of exclusion limit 
close to "Typical experiment" approach.}
    \label{fig:4} 
  \end{center}
\end{figure}

In order to estimate 
the various approaches of the exclusion limit determination 
we suppose that new physics exists, i.e. the value $<N_s>$ equals 
to one of the exclusion limits from Fig.4 
and the value $<N_b>$ equals to the corresponding value of expected background.
Then we apply the equal probability test to find critical value $n_0$
for hypotheses testing in future measurements. Here a zero hypothesis is 
the statement that new physics exists and an alternative hypothesis is the
statement that new physics is absent.
After calculation of the Type I error
$\alpha$ (the probability that the number of observed events will be equal to
or less than the critical value $n_0$)
and the Type II error $\beta$ (the probability that the number of 
observed events will be greater than the critical value $n_0$
in the case of absence of new physics)
we can compare the methods. 
In Table 9 the comparison result is shown. As is seen
from this Table the "Typical experiment" approach~\cite{10} gives
too small values of exclusion limit. The difference in the 90\%~CL definition
is the main reason of the difference between our result and the 
exclusion limit from ref.~\cite{9}. We require that $\epsilon = \kappa$.
In ref~\cite{9} the criterion for determination exclusion limits:
$\beta < \Delta$ and $\displaystyle \frac{\alpha}{1-\beta} < \epsilon$ 
is used, 
i.e. the experiment will observe with probability at least $1 - \Delta$ 
at most a number of events such that the limit obtained at the 
$1 - \epsilon$ confidence level excludes the corresponding 
signal~\footnote{If we define $\epsilon$ as normalized $\kappa$ 
($\displaystyle \epsilon = \tilde \kappa = \frac{\kappa}{2-\kappa}$) 
we have the result close to ref.~\cite{9}.
For example, $\kappa = 0.17$ corresponds to $\epsilon = 0.0929$,
i.e. $1-\epsilon \approx 0.9$.}.

\section{The probability of new physics discovery}

In section 2 we determined the probability $\kappa$ that "new physics" can 
be described by the "standard physics". But it is also very important 
to determine  the probabilty of new physics discovery 
in future experiment. 
According to common definition \cite{1} the new physics discovery corresponds 
to the case when the probability that background can imitate signal is 
less than $5 \sigma$ or in terms of the probabilty less than  
$5.7 \cdot 10^{-7}$ (here of course we neglect any possible systematical 
errors). 

So we require that the probability $\beta(\Delta)$ 
of the background fluctuations for 
$n > n(\Delta)$ is 
less than $\Delta$, namely 
\begin{equation}
\beta(\Delta) =  \sum ^{\infty}_{n=n_0({\Delta})+1} P(<N_b>, n) \leq \Delta
\end{equation} 
The probability $1 - \alpha(\Delta)$
that the number of signal events will be bigger than 
$n_0(\Delta)$ is equal to 
\begin{equation}
1 - \alpha(\Delta) = \sum ^{\infty}_{n = n_0(\Delta) + 1}
P(<N_b> + <N_s>, n)
\end{equation}

It should be stressed that $\Delta$ is a given number and $\alpha(\Delta)$ 
is a function of $\Delta$. Usually physicists claim the 
discovery of phenomenon \cite{1} if the probability of the background 
fluctuation is less than  $5\sigma$ that corresponds to 
$\Delta_{dis} = 5.7 \cdot 10^{-7}$~\footnote{The approximation of 
Poisson distribution by Gaussian for tails with area close to or less than
$\Delta_{dis}$ is wrong.}. 
So from the equation (18) we find 
$n_0(\Delta)$ and estimate the probabilty  $1 - \alpha(\Delta)$ that 
an experiment will satisfy the discovery criterion. 

As an example consider the search for standard Higgs boson 
with a mass $m_h = 110~GeV$ at the CMS 
detector. For total luminosity $ L = 3 \cdot 10^{4} pb^{-1} ( 2 \cdot 
10^{4} pb^{-1})$ one can find \cite{1} that 
$<N_b> = 2893(1929), <N_s> = 357(238)$,
$\displaystyle S_1 = \frac{<N_s>}{\sqrt{<N_b>}} =6.6(5.4)$.
Using the formulae (18, 19) for $\Delta _{dis} = 5.7 \cdot 10^{-7}$ 
($5 \sigma$ discovery criterion) we find that 
$1 - \alpha(\Delta_{dis}) = 0.96(0.73)$. It means that for total 
luminosity $L = 3\cdot 10^{4}pb^{-1}(2\cdot 10^{4}pb^{-1})$ the CMS experiment 
will discover at $\geq 5\sigma$ level 
standard Higgs boson with a mass $m_h~=~110~GeV$ with a probabilty 96(73) 
percent. 

An account of uncertainties related to nonexact knowledge of background
cross section is straightforward and it is based on the results of Section 3.
Suppose uncertainty in the calculation of exact cross section is
determined by parameter $\delta$, i.e. the exact cross section
lies in the interval $(\sigma_b, \sigma_b (1+\delta))$ and
the exact value of average number of events
lies in the interval $(<N_b>, <N_b> (1+\delta))$.
Taking into account formulae (18) and (19)
we have the formulae

\begin{equation}
\beta(\Delta) =  \sum ^{\infty}_{n=n_0({\Delta})+1} P(<N_b>(1+\delta), n) \leq \Delta
\end{equation} 

\begin{equation}
1 - \alpha(\Delta) = \sum ^{\infty}_{n = n_0(\Delta) + 1}
P(<N_b> + <N_s>, n)
\end{equation}

As an application of formulae (20,21) consider the case
$<N_s> = <N_b>= 100$ (typical case for the search for 
supersymmetry at LHC). For such values of $<N_s>$ and $<N_b>$
we have $S_1=10$, $S_2=7.1$, $S_{12}=4.1$. For $\delta$= 0.,
0.1, 0.25, 0.5 we find that $1-\alpha(\Delta_{dis})$ = 0.9998,
0.9938, 0.8793, 0.1696, correspondingly. So, we see that the uncertainty
in the calculations of background cross section is extremely essential for the
determination of the LHC discovery potential. Some other examples
are presented in Tables 10-15.

Let us consider the random variable 
``luminosity of $5\sigma$ discovery claim'' 
for predicted phenomenon in future experiment.
Fig.5 illustrates the behaviour of this value for above example
$<N_s>=<N_b>=100$ at integrated luminosity $10^5pb^{-1}$.
As follow from Fig.5(b) we can point out average luminosity of 
$5\sigma$ discovery claim $\bar L~=~0.3287\cdot10^{5}pb^{-1}$
and estimate the accuracy of this prediction. As seems it is very important
parameter for comparison of proposals of future experiments.

\begin{figure}[htpb]
  \begin{center}
    \resizebox{7cm}{!}{\includegraphics{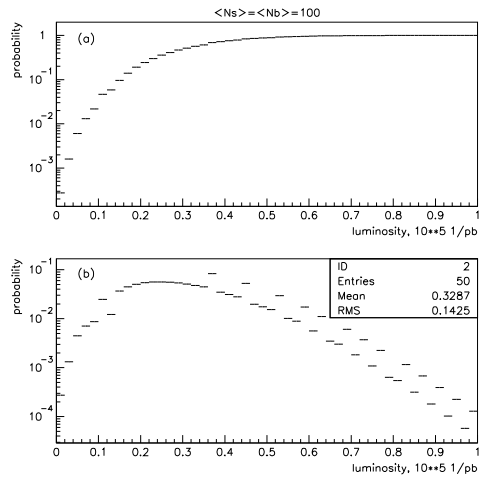}} 
\caption{The cumulative distribution function (a) and 
the behaviour of the probability
distribution (b) of the random variable 
``luminosity of $5\sigma$ discovery claim'' ($<N_s>=<N_b>=100$
 at integrated luminosity $10^5pb^{-1}$).}
    \label{fig:5} 
  \end{center}
\end{figure}
        
\section{Conclusions}

In this paper we determined the probability to discover the new physics 
in the future experiments when the average number of background $<N_b>$ and 
signal events  $<N_s>$ is known. We have found that in this case 
the role of significance plays  $S_{12}~=~\sqrt{<N_b>+<N_s>}~-~\sqrt{<N_b>}$ 
in comparison with often used expressions for the significances
 $S_1~=~\displaystyle~\frac{<N_s>}{\sqrt{<N_b>}}$ and
 $S_2~=~\displaystyle~\frac{<N_s>}{\sqrt{<N_s>~+~<N_b>}}$.

For $<N_s>~\ll~<N_b>$ we have found that $S_{12}  = 0.5 S_1 = 0.5 S_2$.
For not too high values of $<N_s>$ and $<N_b>$, when the deviations
from the Gaussian distributions are essential, our results are presented
in Tables 1-6. We proposed a method for taking into account
systematical errors related to the nonexact knowledge of background and signal
events. An account of such kind of systematics is very essential in the search 
for supersymmetry and leads to an essential decrease in the probability to 
discover the new physics in the future experiments.
We also proposed methods for the estimation of exclusion limits on new
physics and the probability of the new physics discovery in future experiments.

We are indebted to M.Dittmar for very hot discussions and useful questions
which were one of the motivations to perform this study. We are grateful
to V.A.Matveev for the interest and useful comments. This work has been
supported by RFFI grant 99-02-16956.

\begin{table}[h]
\small
    \caption{ The dependence of $\kappa$ on $<N_s>$ and $<N_b>$ for 
$S_1 =5$} 
    \label{tab:Tab.1}
    \begin{center}
\begin{tabular}{|r|r|r|}
\hline
$<N_s>$ & $<N_b>$ & $\kappa$\\ 
\hline
   5 &  1   &  $0.1423$ \\
  10 &  4   &  $0.0828$ \\
  15 &  9   &  $0.0564$ \\
  20 & 16   &  $0.0448$ \\
  25 & 25   &  $0.0383$ \\
  30 & 36   &  $0.0333$ \\
  35 & 49   &  $0.0303$ \\
  40 & 64   &  $0.0278$ \\
  45 & 81   &  $0.0260$ \\
  50 & 100  &  $0.0245$ \\
  55 & 121  &  $0.0234$ \\
  60 & 144  &  $0.0224$ \\
  65 & 169  &  $0.0216$ \\
  70 & 196  &  $0.0209$ \\
  75 & 225  &  $0.0203$ \\
  80 & 256  &  $0.0198$ \\
  85 & 289  &  $0.0193$ \\
  90 & 324  &  $0.0189$ \\
  95 & 361  &  $0.0185$ \\
 100 & 400  &  $0.0182$ \\
 150 & 900  &  $0.0162$ \\
 500 & $10^4$ &  $0.0135$ \\
 5000 & $10^6$ &  $0.0125$ \\
\hline
\end{tabular}
    \end{center}
\end{table}

\begin{table}[h]
\small
    \caption{ The dependence of $\kappa$ on $<N_s>$ and $<N_b>$ 
for $S_2 \approx 5$.}
    \label{tab:Tab.2}
    \begin{center}
\begin{tabular}{|l|l|l|}
\hline
$<N_s>$ & $<N_b>$ & $\kappa$ \\ 
\hline
  26 &   1 &    $0.15 \cdot 10^{-4}$ \\
  29 &   4 &    $0.14 \cdot 10^{-3}$ \\
  33 &   9 &    $0.44 \cdot 10^{-3}$ \\
  37 &  16 &    $0.99 \cdot 10^{-3}$ \\
  41 &  25 &    $0.17 \cdot 10^{-2}$ \\
  45 &  36 &    $0.26 \cdot 10^{-2}$ \\
  50 &  49 &    $0.31 \cdot 10^{-2}$ \\
  55 &  64 &    $0.36 \cdot 10^{-2}$ \\
 100 & 300 &    $0.74 \cdot 10^{-2}$ \\
 150 & 750 &    $0.89 \cdot 10^{-2}$ \\
\hline
\end{tabular}
    \end{center}
\end{table}

\begin{table}[h]
\small
    \caption{$<N_s> = \displaystyle {1 \over 5} \cdot <N_b>$.
The dependence of $\kappa$ on $<N_s>$ and $<N_b>$.}
    \label{tab:Tab.3}
    \begin{center}
\begin{tabular}{|l|l|l|l|l|l|}
\hline
$<N_s>$ & $<N_b>$ & $\kappa$\\ 
\hline
   50&  250&   $0.131$ \\
  100&  500&   $0.033$ \\
  150&  750&   $0.89 \cdot 10^{-2}$ \\
  200& 1000&   $0.25 \cdot 10^{-2}$ \\
  250& 1250&   $0.74 \cdot 10^{-3}$ \\
  300& 1500&   $0.22 \cdot 10^{-3}$ \\
  350& 1750&   $0.65 \cdot 10^{-4}$ \\
  400& 2000&   $0.20 \cdot 10^{-4}$ \\
\hline
\end{tabular}
    \end{center}
\end{table}

\begin{table}[h]
\small
    \caption{$<N_s> = \displaystyle {1 \over 10} \cdot <N_b>$. 
The dependence of $\kappa$ on $<N_s>$ and $<N_b>$.}
    \label{tab:Tab.4}
    \begin{center}
\begin{tabular}{|l|l|l|}
\hline
$<N_s>$ & $<N_b>$ &  $\kappa$\\ 
\hline
   50&  500&   $0.275$ \\
  100& 1000&   $0.123$ \\
  150& 1500&   $0.059$ \\
  200& 2000&   $0.029$ \\
  250& 2500&   $0.015$ \\
  300& 3000&   $0.75 \cdot 10^{-2}$ \\
  350& 3500&   $0.38 \cdot 10^{-2}$ \\
  400& 4000&   $0.20 \cdot 10^{-2}$ \\
  450& 4500&   $0.11 \cdot 10^{-2}$ \\
  500& 5000&   $0.56 \cdot 10^{-3}$ \\
\hline
\end{tabular}
    \end{center}
\end{table}

\begin{table}[h]
\small
    \caption{$<N_s> = <N_b>$. 
The dependence of $\kappa$ on $<N_s>$ and $<N_b>$.}
    \label{tab:Tab.5}
    \begin{center}
\begin{tabular}{|l|l|l|}
\hline
$<N_s>$ & $<N_b>$ &  $\kappa$\\ 
\hline
    2. &   2. &     0.561    \\
    4. &   4. &     0.406   \\
    6. &   6. &     0.308   \\
    8. &   8. &     0.239    \\
   10. &  10. &     0.188  \\
   12. &  12. &     0.150  \\
   14. &  14. &     0.121    \\
   16. &  16. &     0.098 \\
   18. &  18. &     0.079 \\
   20. &  20. &     0.064 \\
   24. &  24. &     0.042 \\
   28. &  28. &     0.028 \\
   32. &  32. &     0.019 \\
   36. &  36. &     0.013 \\
   40. &  40. &    $0.87 \cdot 10^{-2}$ \\
   50. &  50. &    $0.34 \cdot 10^{-2}$ \\
   60. &  60. &    $0.13 \cdot 10^{-2}$ \\
   70. &  70. &    $0.52 \cdot 10^{-3}$ \\
   80. &  80. &    $0.21 \cdot 10^{-3}$ \\
  100. & 100. &    $0.33 \cdot 10^{-4}$ \\
\hline
\end{tabular}
    \end{center}
\end{table}

\begin{table}[h]
\small
    \caption{$<N_s> = 2 \cdot <N_b>$. 
The dependence of $\kappa$ on $<N_s>$ and $<N_b>$.}
    \label{tab:Tab.6}
    \begin{center}
\begin{tabular}{|l|l|l|}
\hline
$<N_s>$ & $<N_b>$ &  $\kappa$\\ 
\hline
    2. &   1. &     0.463    \\
    4. &   2. &     0.294   \\
    6. &   3. &     0.200   \\
    8. &   4. &     0.141    \\
   10. &   5. &     0.102  \\
   12. &   6. &     0.073  \\
   14. &   7. &     0.052    \\
   16. &   8. &     0.037 \\
   18. &   9. &     0.027 \\
   20. &  10. &     0.020 \\
   24. &  12. &     0.011 \\
   28. &  14. &    $0.59 \cdot 10^{-2}$ \\
   32. &  16. &    $0.33 \cdot 10^{-2}$ \\
   36. &  18. &    $0.18 \cdot 10^{-2}$ \\
   40. &  20. &    $0.10 \cdot 10^{-2}$ \\
   50. &  25. &    $0.23 \cdot 10^{-3}$ \\
   60. &  30. &    $0.56 \cdot 10^{-4}$ \\
\hline
\end{tabular}
    \end{center}
\end{table}

\begin{table}[h]
\small
    \caption{$90\%$ exclusion limits on signal cross section for
$L = 10^4pb^{-1}$ and for different background cross section
(everything in pb). The third column gives exclusion limit
according to formula (15).}
    \label{tab:Tab.7}
    \begin{center}
\begin{tabular}{|r|r|r|}
\hline
$\sigma_b$ & $\sigma_s$ &  $\sigma_s$ (continuous limit)\\ 
\hline
$10^3$   & $1.041$ & 1.038 \\
$10^2$   & $0.329$ & 0.328 \\
$10$     & $0.104$ & 0.104 \\
$1$      & $0.033$ & 0.033 \\
$0.1$    & $0.011$ & 0.011 \\
$0.01$   & $0.0036$ & 0.0035 \\
$0.001$  & $0.0013$ & 0.0013 \\
$0.0001$ & $0.00060$ & 0.00060 \\
\hline
\end{tabular}
    \end{center}
\end{table}

\begin{table}[h]
\small
    \caption{$90\%$ exclusion limits on signal cross section for
$L = 10^4pb^{-1}$, $2 \delta_{1b} = 0.25$ 
and for different background cross section
(everything in pb). The third column gives exclusion limit
according to formula (17).}
    \label{tab:Tab.8}
    \begin{center}
\begin{tabular}{|r|r|r|}
\hline
$\sigma_b$ & $\sigma_s$ &  $\sigma_s$ (continuous limit)\\ 
\hline
  $10^3$ & $ 251.25 $ &251.16 \\
  $10^2$ & $  25.37 $ & 25.37 \\
    $10$ & $ 2.62 $ &  2.62 \\
     $1$ & $ 0.29 $ &  0.29 \\
   $0.1$ & $ 0.037 $ &  0.037 \\
  $0.01$ & $ 0.0064 $ &  0.0064 \\
 $0.001$ & $ 0.0017 $ &  0.0017 \\
$0.0001$ & $ 0.00064 $ &  0.00066 \\
\hline
\end{tabular}
    \end{center}
\end{table}

\begin{table}
    \begin{center}
    \caption{The comparison of the different approaches to determination
of the exclusion limits. The $\alpha$ and the $\beta$ are the Type I and 
the Type II errors under the equal probability test. The $\kappa$ equals to
the sum of $\alpha$ and $\beta$.}
    \label{tab:Tab.9}
\begin{tabular}{|r|rrrr|rrrr|rrrr|}
\hline
     &  &this&paper & & & ref.& [9] & & &ref.&[10]&  \\ 
\hline
$N_b$&$N_s$&$\alpha$&$\beta$&$\kappa$&$N_s$&$\alpha$&$\beta$&$\kappa$
&$N_s$&$\alpha$&$\beta$&$\kappa$\\ 
\hline
  1& 6.02&0.08 &0.02 &0.10 & 4.45 &0.09 &0.08 &0.17  &3.30 &0.20 &0.08 &0.28\\
  2& 7.25&0.05 &0.05 &0.10 & 5.50 &0.13 &0.05 &0.18  &3.90 &0.16 &0.14 &0.30\\
  3& 8.32&0.07 &0.03 &0.10 & 6.40 &0.09 &0.08 &0.18  &4.40 &0.14 &0.18 &0.32\\
  4& 9.20&0.05 &0.05 &0.10 & 7.25 &0.13 &0.05 &0.18  &4.80 &0.23 &0.11 &0.34\\
  5&10.06&0.07 &0.03 &0.10 & 7.90 &0.10 &0.07 &0.17  &5.20 &0.20 &0.13 &0.34\\
  6&10.67&0.06 &0.04 &0.10 & 8.41 &0.09 &0.08 &0.18  &5.50 &0.19 &0.15 &0.34\\
  7&11.37&0.05 &0.05 &0.10 & 9.00 &0.08 &0.10 &0.18  &5.90 &0.17 &0.17 &0.34\\
  8&12.02&0.07 &0.03 &0.10 & 9.70 &0.10 &0.06 &0.17  &6.10 &0.17 &0.18 &0.35\\
  9&12.51&0.06 &0.04 &0.10 &10.16 &0.09 &0.07 &0.17  &6.40 &0.16 &0.20 &0.36\\
 10&13.04&0.05 &0.05 &0.10 &10.50 &0.09 &0.08 &0.17  &6.70 &0.22 &0.14 &0.36\\
 11&13.62&0.04 &0.06 &0.10 &10.80 &0.08 &0.09 &0.18  &6.90 &0.21 &0.15 &0.36\\
\hline
\end{tabular}
    \end{center}
\end{table}

\begin{table}[h]
    \caption{The dependence of $1-\alpha(\Delta_{dis})$ on
$<N_s>$ and $<N_b>$ for $S_1=5$ and different values of $\delta$.}
    \label{tab:Tab.10}
    \begin{center}
\begin{tabular}{|r|r|l|l|l|l|}
\hline
$<N_s>$&$<N_b>$&$\delta=0.0$&$\delta=0.1$&$\delta=0.25$&$\delta=0.5$\\ 
\hline
   5  &    1 &   0.0839  &     0.0839  &     0.0426   &    0.0426\\
  10  &    4 &   0.1728  &     0.1174  &     0.0765   &    0.0288\\
  15  &    9 &   0.2323  &     0.1321  &     0.0678   &    0.0132\\
  20  &   16 &   0.2737  &     0.1783  &     0.0609   &    0.0071\\
  25  &   25 &   0.3041  &     0.1779  &     0.0424   &    0.0020\\
  30  &   36 &   0.3273  &     0.1480  &     0.0315   &    0.0007\\
  35  &   49 &   0.3456  &     0.1502  &     0.0192   &    0.0001\\
  40  &   64 &   0.3973  &     0.1305  &     0.0125   &    0.00003\\
  45  &   81 &   0.4064  &     0.1157  &     0.0068   &    0.000004\\
  50  &  100 &   0.4140  &     0.1042  &     0.0040   &  \\
  55  &  121 &   0.4205  &     0.0950  &     0.0019   &  \\
  60  &  144 &   0.4261 &      0.0876  &     0.0010   &  \\
  65  &  169 &   0.4309  &     0.0723  &     0.0004   &  \\
  70  &  196 &   0.4352  &     0.0606  &     0.0002   &  \\
  75  &  225 &   0.4389  &     0.0516  &     0.0001   &  \\
  80  &  256 &   0.4638  &     0.0444  &     0.00003  &  \\
  85  &  289 &   0.4657  &     0.0387  &     0.00001  &  \\
  90  &  324 &   0.4674  &     0.0306  &     0.000005 &  \\
  95  &  361 &   0.4689  &     0.0245  &     0.000002 &  \\
 100  &  400 &   0.4703  &     0.0199  &              &  \\
 150  &  900 &   0.5041  &     0.0015  &              &  \\
\hline
\end{tabular}
    \end{center}
\end{table}

\begin{table}[h]
    \caption{The dependence of $1-\alpha(\Delta_{dis})$ on
$<N_s>$ and $<N_b>$ for $S_2 \approx 5$ and different values of $\delta$.}
    \label{tab:Tab.11}
    \begin{center}
\begin{tabular}{|r|r|l|l|l|l|}
\hline
$<N_s>$&$<N_b>$&$\delta=0.$&$\delta=0.1$&$\delta=0.25$&$\delta=0.5$\\ 
\hline
  26 &     1 &   0.9999 &      0.9999  &     0.9998  &     0.9998    \\
  29 &     4 &   0.9983 &      0.9968  &     0.9940  &     0.9825    \\
  33 &     9 &   0.9909 &      0.9779  &     0.9524  &     0.8423    \\
  37 &    16 &   0.9725 &      0.9473  &     0.8491  &     0.5730    \\
  41 &    25 &   0.9418 &      0.8806  &     0.6606  &     0.2457    \\
  45 &    36 &   0.9016 &      0.7622  &     0.4705  &     0.0848\\
  50 &    49 &   0.8774 &      0.7058  &     0.3208  &     0.0222\\
  55 &    64 &   0.8752 &      0.6206  &     0.2161  &     0.0057\\
 100 &   300 &   0.7155 &      0.1307  &     0.0002  & \\
 150 &   750 &   0.6599 &      0.0119  &             &\\
\hline
\end{tabular}
    \end{center}
\end{table}

\begin{table}[h]
  \caption{$<N_s> = \frac{1}{5} \cdot <N_b>$. The dependence of $1-\alpha(\Delta_{dis})$ 
on $<N_s>$ and $<N_b>$ for different values of~$\delta$.}
    \label{tab:Tab.12}
    \begin{center}
\begin{tabular}{|r|r|l|l|}
\hline
$<N_s>$&$<N_b>$&$\delta=0.$&$\delta=0.1$\\ 
\hline
     50 &   250  &  0.0408  &     0.0004  \\ 
    100 &   500  &  0.3032  &     0.0030  \\ 
    150 &   750  &  0.6599  &     0.0119  \\ 
    200 &  1000  &  0.8905  &     0.0301  \\ 
    250 &  1250  &  0.9735  &     0.0629  \\ 
    300 &  1500  &  0.9947  &     0.1127  \\     
    350 &  1750  &  0.9992  &     0.1767  \\    
    400 &  2000  &  0.9999  &     0.2595  \\     
\hline
\end{tabular}
    \end{center}
\end{table}

\begin{table}[h]
  \caption{$<N_s> = \frac{1}{10} \cdot <N_b>$. The dependence 
of $1-\alpha(\Delta_{dis})$ on $<N_s>$ and $<N_b>$.}
    \label{tab:Tab.13}
    \begin{center}
\begin{tabular}{|r|r|l|}
\hline
$<N_s>$&$<N_b>$&$\delta=0.$\\ 
\hline
     50 &   500 &   0.0043  \\
    100 &  1000 &   0.0424  \\
    150 &  1500 &   0.1478  \\    
    200 &  2000 &   0.3223  \\    
    250 &  2500 &   0.5177  \\    
    300 &  3000 &   0.6955  \\    
    350 &  3500 &   0.8270  \\    
    400 &  4000 &   0.9093  \\    
\hline
\end{tabular}
    \end{center}
\end{table}

\begin{table}[h]
    \caption{$<N_s>$ = $<N_b>$. The dependence of $1-\alpha(\Delta_{dis})$ on
$<N_s>$ and $<N_b>$ for different values of $\delta$.}
    \label{tab:Tab.14}
    \begin{center}
\begin{tabular}{|r|r|l|l|l|l|}
\hline
$<N_s>$&$<N_b>$&$\delta=0.$&$\delta=0.1$&$\delta=0.25$&$\delta=0.5$\\ 
\hline
      2  &    2 &   0.0003    &   0.0003    &   0.0001    &   0.000005\\
      4  &    4 &   0.0016    &   0.0007    &   0.0003    &   0.00003\\
      6  &    6 &   0.0061    &   0.0030    &   0.0007    &   0.00006\\
      8  &    8 &   0.0131   &    0.0041    &   0.0011    &   0.0001\\
     10  &   10 &   0.0218    &   0.0081    &   0.0027    &   0.0002\\
     12  &   12 &   0.0467    &   0.0206    &   0.0050    &   0.0003\\
     14  &   14 &   0.0589    &   0.0283    &   0.0080    &   0.0004\\
     16  &   16 &   0.0956    &   0.0512    &   0.0116    &   0.0007\\
     18  &   18 &   0.1401    &   0.0609    &   0.0156    &   0.0007\\
     20  &   20 &   0.1903    &   0.0925    &   0.0200    &   0.0012\\
     24  &   24 &   0.3005    &   0.1402    &   0.0395    &   0.0017\\
     28  &   28 &   0.4122    &   0.2280    &   0.0656    &   0.0031\\
     32  &   32 &   0.5166    &   0.2821    &   0.0969    &   0.0050\\
     36  &   36 &   0.6089    &   0.3773    &   0.1323    &   0.0073\\
     40  &   40 &   0.7268    &   0.4703    &   0.1704    &   0.0101\\
     50  &   50 &   0.8762    &   0.6688    &   0.3216    &   0.0181\\
     60  &   60 &   0.9572    &   0.8309    &   0.4397    &   0.0332\\
     70  &   70 &   0.9865    &   0.9206    &   0.5784    &   0.0612\\
     80  &   80 &   0.9960    &   0.9648    &   0.7205    &   0.0850\\
    100  &  100 &   0.9998    &   0.9938    &   0.8793    &   0.1696  \\  
\hline
\end{tabular}
    \end{center}
\end{table}

\begin{table}[h]
  \caption{$<N_s>~=~0.5 \cdot <N_b>$. The dependence of 
$1-\alpha(\Delta_{dis})$ on $<N_s>$ and $<N_b>$ for different 
values of $\delta$.}
    \label{tab:Tab.15}
    \begin{center}
\begin{tabular}{|r|r|l|l|l|}
\hline
$<N_s>$&$<N_b>$&$\delta=0.$&$\delta=0.1$&$\delta=0.25$\\ 
\hline
      2 &     4 &   0.0001 &      0.00002 &     0.000005 \\
      4 &     8 &   0.0003 &      0.0001 &      0.000009 \\
      6 &    12 &   0.0010 &      0.0002 &      0.00003 \\
      8 &    16 &   0.0017 &      0.0005 &      0.00004 \\
     10 &    20 &   0.0040 &      0.0009 &      0.00005 \\
     12 &    24 &   0.0071 &      0.0012 &      0.0001 \\
     14 &    28 &   0.0111 &      0.0023 &      0.0001 \\
     16 &    32 &   0.0156 &      0.0025 &      0.0002\\
     18 &    36 &   0.0207 &      0.0039 &      0.0002 \\
     20 &    40 &   0.0341 &      0.0056 &      0.0003 \\
     24 &    48 &   0.0589 &      0.0099 &      0.0005 \\
     28 &    56 &   0.0886 &      0.0192 &      0.0008\\
     32 &    64 &   0.1424 &      0.0259 &      0.0011 \\
     36 &    72 &   0.1796 &      0.0402 &      0.0013 \\
     40 &    80 &   0.2442 &      0.0575 &      0.0021 \\
     50 &   100 &   0.4140 &      0.1042 &      0.0040 \\
     60 &   120 &   0.5692 &      0.1947 &      0.0074 \\
     70 &   140 &   0.7187 &      0.2762 &      0.0118 \\
     80 &   160 &   0.8250 &      0.3820 &      0.0195 \\
    100 &   200 &   0.9456 &      0.5765 &      0.0408 \\
\hline
\end{tabular}
    \end{center}
\end{table}

\end{document}